\documentclass[12pt]{iopart}
\usepackage{graphicx}

\def\PRL{{\em Phys. Rev. Lett.}}
\def\PRC{{\em Phys. Rev.} C}
\def\PRD{{\em Phys. Rev.} D}
\def\PLB{{\em Phys. Lett.} B}
\def\NPA{{\em Nucl. Phys.} A}
\def\NPB{{\em Nucl. Phys.} B}
\def\IJMPE{{\em Int. J. Mod. Phys.} E}
\def\EPJA{{\em Eur. Phys. J.} A}
\def\AP{{\em Ann. Phys. (N.Y.)}}
\def\AR{{\em Annu. Rev. Nucl. Part. Sci.}}
\def\Journal#1#2#3#4{#4 #1 {\bf #2} #3}

\newcommand{\ba}{\begin{eqnarray}}
\newcommand{\ea}{\end{eqnarray}}

\begin{document}
\pagestyle{plain}

\letter{Strange form factors of the proton in a two-component model}  

\author{R. Bijker \footnote{E-mail: bijker@nucleares.unam.mx}}

\address{Instituto de Ciencias Nucleares, 
Universidad Nacional Aut\'onoma de M\'exico, 
AP 70-543, 04510 M\'exico, D.F., M\'exico}

\begin{abstract}
The strange form factors of the nucleon are studied in a two-component 
model consisting of a three-quark intrinsic structure surrounded by a 
meson cloud. A comparison with the available experimental world data from the 
SAMPLE, PVA4, HAPPEX and G0 collaborations shows a good overall agreement. 
The strange magnetic moment is found to be positive $\mu_s=0.315$ $\mu_N$. 
\end{abstract}

\pacs{13.40.Gp, 12.40.Vv, 14.20.Dh, 24.85.+p, 13.40.Em}

\submitto{\JPG}

\maketitle                   

\section{Introduction}

Electromagnetic and weak form factors are key ingredients to the 
understanding of the internal structure of the nucleon, since they 
contain the information about the distributions of electric charge and 
magnetization. Deep-inelastic scattering experiments have demonstrated 
that the structure of the proton cannot be described by its $uud$ 
valence structure alone: the valence quarks carry less than 50 \% of 
the proton momentum and less than 30 \% of the proton spin. 
Especially, the contribution of strange quarks to the nucleon structure 
is of interest because it is exclusively part of the quark-antiquark 
sea. In recent experiments, parity-violating elastic electron-proton 
scattering has been used to probe the contribution of strange quarks 
to the structure of the nucleon. The strange quark content of the 
form factors is determined assuming charge symmetry and combining 
parity-violating asymmetries with measurements of the electric and 
magnetic form factors of the proton and neutron \cite{Manohar}. 

The first experimental results indicate that the strangeness content 
of both the magnetic moment and the radius of the proton is positive 
\cite{Aniol05b}, an unexpected and surprising finding, since a majority 
of theoretical studies favors a negative value for both quantities 
\cite{beck}. The aim of this Letter is to present a study of strange 
form factors in a two-component model of the nucleon \cite{IJL,BI} and 
to analyze the available experimental data. 

\section{Two-component model}

The momentum dependence of the current matrix elements is contained 
in the Dirac and Pauli form factors, $F_{1}(Q^2)$ and $F_2(Q^2)$, 
respectively. The electric and magnetic (Sachs) form factors are 
obtained from $F_{1}$ and $F_{2}$ by the relations $G_E=F_1-\tau F_2$ 
and $G_M=F_1 + F_2$ with $\tau=Q^2/4 M_N^2$. 
The Dirac and Pauli form factors are parametrized according to a  
two-component model of the nucleon \cite{BI} in which the external photon 
couples both to an intrinsic three-quark structure described by the form 
factor $g(Q^2)$ and to a meson cloud through the intermediate vector mesons  
$\rho$, $\omega$ and $\phi$. In the original version of the two-component model 
\cite{IJL}, the Dirac form factor was attributed to both the intrinsic structure and 
the meson cloud, and the Pauli form factor entirely to the meson cloud. 
In a modified version \cite{BI}, it was shown that the addition of an intrinsic 
part to the isovector Pauli form factor as suggested by studies of 
relativistic constituent quark models in the light-front approach 
\cite{frank}, improves the results for the elecromagnetic form factors 
of the neutron considerably. 

The isoscalar and isovector form factors correspond to the currents  
$J_{\mu}^{I=0}=(\bar{u} \gamma_{\mu} u + \bar{d} \gamma_{\mu} d 
-2 \bar{s} \gamma_{\mu} s)/6$ 
and $J_{\mu}^{I=1}=(\bar{u} \gamma_{\mu} u - \bar{d} \gamma_{\mu} d)/2$, respectively. 
The isoscalar Dirac and Pauli form factors 
contain the couplings to the $\omega$ and $\phi$ mesons 
\ba
F_{1}^{I=0}(Q^{2}) &=& \frac{1}{2} g(Q^{2}) \left[ 
1-\beta_{\omega}-\beta_{\phi}
+\beta_{\omega} \frac{m_{\omega }^{2}}{m_{\omega }^{2}+Q^{2}} 
+\beta_{\phi} \frac{m_{\phi}^{2}}{m_{\phi }^{2}+Q^{2}}\right] , 
\nonumber\\
F_{2}^{I=0}(Q^{2}) &=& \frac{1}{2}g(Q^{2})\left[ 
\alpha_{\omega} \frac{m_{\omega }^{2}}{m_{\omega }^{2}+Q^{2}} 
+ \alpha_{\phi} \frac{m_{\phi}^{2}}{m_{\phi}^{2}+Q^{2}}\right] ,
\ea
and the isovector ones the coupling to the $\rho$ meson \cite{BI}
\ba 
F_{1}^{I=1}(Q^{2}) &=& \frac{1}{2}g(Q^{2})\left[ 1-\beta_{\rho} 
+\beta_{\rho} \frac{m_{\rho}^{2}}{m_{\rho}^{2}+Q^{2}} \right] , 
\nonumber\\ 
F_{2}^{I=1}(Q^{2}) &=& \frac{1}{2}g(Q^{2})\left[ 
\frac{\mu_{p}-\mu_{n}-1-\alpha_{\rho}}{1+\gamma Q^{2}} 
+\alpha_{\rho} \frac{m_{\rho }^{2}}{m_{\rho}^{2}+Q^{2}} \right] .
\ea
This parametrization ensures that the three-quark contribution to the 
anomalous magnetic moment is purely isovector, as given by $SU(6)$. 
The intrinsic form factor is a dipole $g(Q^{2})=(1+\gamma Q^{2})^{-2}$ 
which coincides with the form used in an algebraic treatment of the 
intrinsic three-quark structure \cite{bijker}. 
The large width of the $\rho$ meson which is crucial for the small $Q^{2}$ 
behavior of the form factors, is taken into account in the same way as in 
\cite{IJL,BI}. For small values of $Q^2$ the form factors are dominated by the 
meson dynamics, whereas for large values they satisfy the asymptotic behavior 
of p-QCD, $F_1 \sim 1/Q^4$ and $F_2 \sim 1/Q^6$ \cite{pQCD}.

\section{Strange form factors}

The strange quark content of the nucleon form factors arises through 
the coupling of the strange current $J_{\mu}^s= \bar{s} \gamma_{\mu} s$ 
to the intermediate vector mesons  $\omega$ and $\phi$. Note that here   
the convention of Jaffe \cite{Jaffe} for the strangeness current has 
been adopted. The wave functions of the $\omega$ and $\phi$ mesons are 
given by
\ba
\left| \omega \right> &=& \cos \epsilon \left| \omega_0 \right> 
- \sin \epsilon \left| \phi_0 \right> ,
\nonumber\\
\left| \phi \right> &=& \sin \epsilon \left| \omega_0 \right> 
+ \cos \epsilon \left| \phi_0 \right> ,
\ea
where the mixing angle $\epsilon$ represents the deviation from the 
ideally mixed states  
$\left| \omega_0 \right>=\left( u \bar{u} + d \bar{d} \right)/\sqrt{2}$ 
and $\left| \phi_0 \right> = s \bar{s}$. 
Under the assumption that the strange form factors have the same form as 
the isoscalar ones, the Dirac and Pauli form factors that correspond to the 
strange current are expressed as the product  
of an intrinsic part $g(Q^2)$ and a contribution from the vector mesons 
\ba
F_{1}^{s}(Q^{2}) &=& \frac{1}{2}g(Q^{2})\left[ 
\beta_{\omega}^s \frac{m_{\omega}^{2}}{m_{\omega }^{2}+Q^{2}} 
+\beta_{\phi}^s \frac{m_{\phi}^{2}}{m_{\phi }^{2}+Q^{2}}\right] , 
\nonumber\\
F_{2}^{s}(Q^{2}) &=& \frac{1}{2}g(Q^{2})\left[ 
\alpha_{\omega}^s \frac{m_{\omega}^{2}}{m_{\omega }^{2}+Q^{2}}
+\alpha_{\phi}^s \frac{m_{\phi}^{2}}{m_{\phi }^{2}+Q^{2}}\right] .
\ea
The coefficients $\beta$ and $\alpha$ appearing in the isoscalar and 
strange form factors depend on the same meson-nucleon and meson-current 
couplings. Following the procedure of \cite{Jaffe}, the vector 
meson-nucleon coupling for the ideally mixed states is parametrized as 
$g_i(\phi_0 N) = g_i \sin \eta_i$ and $g_i(\omega_0 N) = g_i \cos \eta_i$ 
for each of the two Dirac couplings ($i=1$ for the vector coupling 
$\gamma_{\mu}$ and $i=2$ for the tensor coupling $\sigma_{\mu\nu}q^{\nu}$), 
and it is assumed that the quark in the 
vector meson only couples to the quark current of the same flavor with 
a flavor-independent strength $\kappa$. As a result, the isoscalar and 
strange couplings depend on four coefficients $\kappa g_1$, $\kappa g_2$, 
$\eta_1$ and $\eta_2$. However, these couplings are constrained by 
the electric charges and magnetic moments of the nucleon 
\ba
\alpha_{\omega} &=& \mu_p + \mu_n -1 - \alpha_{\phi} ,
\nonumber\\
\beta_{\omega}^s &=& - \beta_{\phi}^s ,
\label{coef1}
\ea
which reduces the number of independent coefficients to two only. 
The latter condition in (\ref{coef1}) is a consequence of the fact that 
the strange (anti)quarks do not contribute to the electric charge,  
$G_{E}^{s}(0)=F_1^s(0)=0$ (leading to $\eta_1=0$).  
The explicit expressions for the strange couplings are 
\ba
\beta_{\omega}^s &=& -\kappa g_1 \sin \epsilon \cos \epsilon , 
\nonumber\\
\beta_{\phi}^s &=& +\kappa g_1 \cos \epsilon \sin \epsilon , 
\nonumber\\
\alpha_{\omega}^s &=& -\kappa g_2 \sin \epsilon \cos (\eta_2 + \epsilon) , 
\nonumber\\
\alpha_{\phi}^s &=& +\kappa g_2 \cos \epsilon \sin (\eta_2 + \epsilon) .  
\label{strange}
\ea
They are related to the isoscalar couplings by \cite{Jaffe} 
\ba
\beta_{\omega}^s/\beta_{\omega} =  
\alpha_{\omega}^s/\alpha_{\omega} &=& 
-\sqrt{6} \, \sin \epsilon/\sin(\theta_0+\epsilon) ,
\nonumber\\
\beta_{\phi}^s/\beta_{\phi} =  
\alpha_{\phi}^s/\alpha_{\phi} &=& 
-\sqrt{6} \, \cos \epsilon/\cos(\theta_0+\epsilon) ,  
\label{coef2}
\ea
where the angle $\theta_0$ is defined by $\tan \theta_0 = 1/\sqrt{2}$. 
Equations (\ref{coef1}) and (\ref{coef2}) imply that the isocalar couplings 
in the Dirac form factor are related to one another by 
\ba
\beta_{\phi} = - \beta_{\omega} \tan \epsilon/\tan(\theta_0+\epsilon) .
\label{coef3}
\ea
This is in contrast with \cite{BI}, where $\beta_{\omega}$ and $\beta_{\phi}$ 
were treated are independent coefficients. 

\section{Results}

\begin{table}
\centering
\caption[]{\small Parameter values obtained in a fit to the 
electromagnetic form factors of the nucleon. $\gamma$ is given in 
terms of (GeV/c)$^{-2}$. The values marked by 
an asterisk are obtained from (\ref{coef1}-\ref{coef3}).}
\label{fit}
\vspace{15pt}
\begin{tabular}{crrcr}
\hline
& Ref.~\cite{BI} & Present & & Eq.~(\ref{coef2}) \\
\hline
$\gamma$ & 0.515 & 0.512 & & \\
$\beta_{\omega}$ & 1.129 & 0.964 & $\beta_{\omega}^s$ & $-0.202^{*}$ \\
$\alpha_{\omega}$ & 0.080$^{*}$ & 0.088$^{*}$ & $\alpha_{\omega}^s$ & $-0.018^{*}$ \\ 
$\beta_{\phi}$ & $-0.263$ & $-0.065^{*}$ & $\beta_{\phi}^s$ & 0.202$^{*}$ \\
$\alpha_{\phi}$ & $-0.200$ & $-0.208$ & $\alpha_{\phi}^s$ & 0.648$^{*}$ \\
$\beta_{\rho}$ & 0.512 & 0.504 & \\
$\alpha_{\rho}$ & 2.675 & 2.705 & \\
\hline
\end{tabular}
\end{table}

In order to calculate the nucleon form factors in the two-component model, the 
five remaining coefficients, $\gamma$ from the intrinsic form factor, $\beta_{\omega}$ 
and $\alpha_{\phi}$ from the isoscalar couplings, and $\beta_{\rho}$ and  
$\alpha_{\rho}$ from the isovector couplings, are determined in a least-square 
fit to the electromagnetic form factors of the proton and the neutron using the 
same data set as in \cite{BI}. The mixing angle $\epsilon$ can be determined either 
from the radiative decays of the $\omega$ and $\phi$ mesons \cite{Jain,Iachello,Harada} 
or from their strong decays \cite{Gobbi}. The value used here is $\epsilon=0.053$ 
rad \cite{Jain}. The values of the five coefficients are very close 
to those obtained in \cite{BI}, as can be seen from Table~\ref{fit}. Note that 
in \cite{BI}, the sum of $\beta_{\omega}$ and $\beta_{\phi}$ could 
be determined well, but their individual values not. Table~\ref{fit} shows that 
indeed the sum of $\beta_{\omega}$ and $\beta_{\phi}$ is almost the same in both 
cases. In the present calculation their ratio is determined by 
the mixing angle $\epsilon$ according to (\ref{coef3}). 
The spatial extent of the intrinsic structure is found to be  
$\langle r^{2}\rangle ^{1/2} = \sqrt{12 \gamma} = 0.49$ fm. 
The last column of Table~\ref{fit} shows the values of the strange couplings as 
obtained from (\ref{coef2}). 

\begin{figure}[htb]
\includegraphics{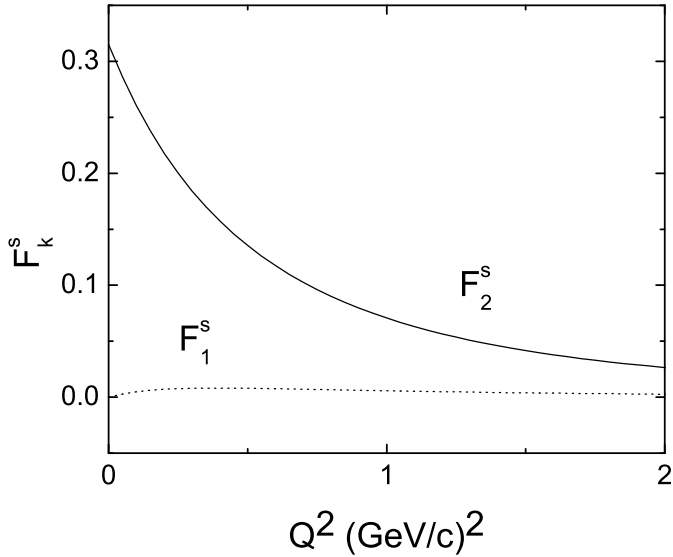} 
\caption[]{\small
Strange Dirac and Pauli form factors, $F_1^s$ (dotted line) and 
$F_2^s$ (solid line).}
\label{f12s}
\end{figure}

Figure~\ref{f12s} shows the strange Dirac and Pauli form factors as a function 
of the momentum transfer $Q^2$.  
Whereas the Pauli form factor is dominated by the coupling to the $\phi$ 
meson ($\alpha_{\phi}^s \gg \alpha_{\omega}^s$), the Dirac form factor is 
small due to a cancelation between the contributions from the $\omega$ and $\phi$ 
mesons ($\beta_{\phi}^s=-\beta_{\omega}^s$). The qualitative features of these 
form factors can be understood in the limit of ideally 
mixed mesons, {\em i.e.} zero mixing angle $\epsilon=0$ (in comparison to 
the value of $\epsilon=3.0^{\circ}$ used in Figure~\ref{f12s}). Since in this case  
$\beta_{\phi}^s=\beta_{\omega}^s=\alpha_{\omega}^s=0$, the Dirac form factor 
vanishes identically and the Pauli form factor depends only on the tensor coupling 
to the $\phi$ meson $\alpha_{\phi}^s$  
\ba
F_1^s(Q^2) &=& 0 ,
\nonumber\\ 
F_2^s(Q^2) &=& \frac{1}{2} g(Q^2) \alpha_{\phi}^s \frac{m_{\phi}^2}{m_{\phi}^2 + Q^2} .
\ea 
The behavior of $F_1^s$ and $F_2^s$ in Figure~\ref{f12s} is quite different from 
that obtained in other theoretical approaches, especially for the strange Pauli 
form factor. Almost all calculations give negative values for $F_2^s$ for the same 
range of $Q^2$ values \cite{Jaffe,Park,Garvey,Forkel,Hammer,Lyubovitskij},  
with the exception of the meson-exchange model \cite{Meissner} 
and the $SU(3)$ chiral quark-soliton model \cite{Silva}. 
In the former case, the values of $F_2^s$ are about two orders of 
magnitude smaller than the present ones, whereas in the latter $F_2^s$  
is positive for small values of $Q^2$, but changes sign around $Q^2=0.1-0.3$ 
(GeV/c)$^2$.

\begin{figure}[htb]
\includegraphics{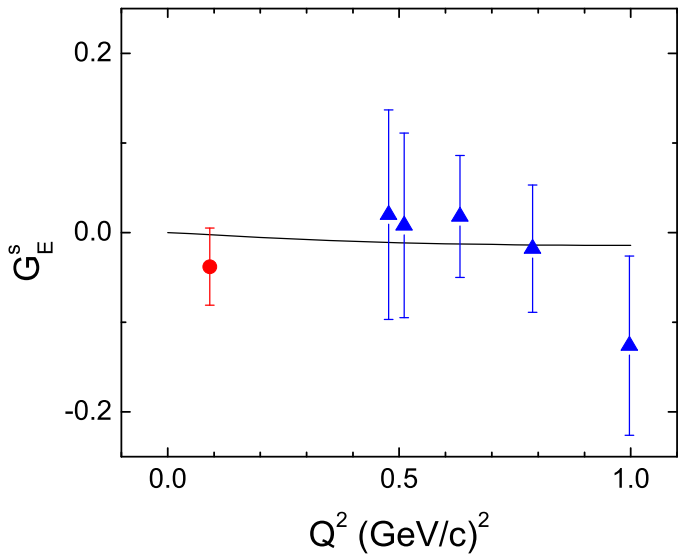} 
\caption[]{\small 
Comparison between theoretical and experimental values of the strange 
electric form factor. The experimental values are taken from \cite{Aniol05a} 
(circle) and \cite{Frascati} (triangle).}
\label{GEs}
\end{figure}

\begin{figure}[htb]
\includegraphics{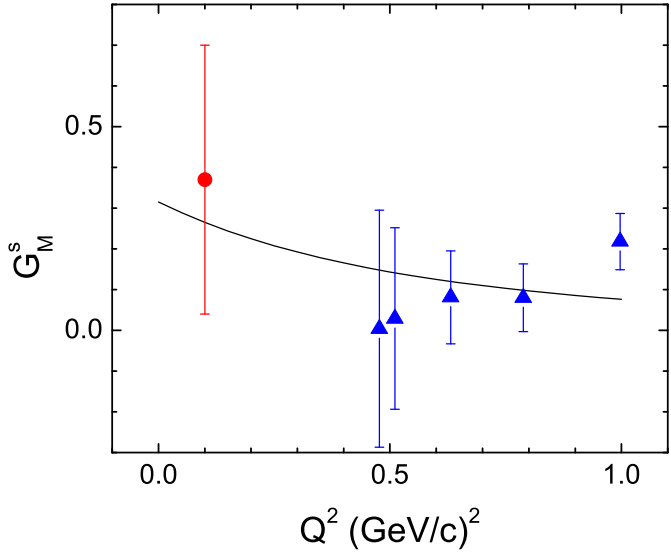} 
\caption[]{\small 
Comparison between theoretical and experimental values of the strange 
magnetic form factor. The experimental values are taken from \cite{Spayde} 
(circle) and \cite{Frascati} (triangle).}
\label{GMs}
\end{figure}

Figures~\ref{GEs} and \ref{GMs} show the strange electric and magnetic form 
factors as a function of $Q^2$. The theoretical values for $G_E^s$ are small 
and negative, in agreement with the recent experimental result of the HAPPEX 
Collaboration in which $G_E^s$ was determined in parity-violating 
electron scattering from $^{4}$He \cite{Aniol05a}. 
The experimental value $G_E^s=-0.038 \pm 0.042 \pm 0.010$ 
measured at $Q^2=0.091$ (GeV/c)$^2$ is consistent with zero. 
The values of $G_M^s$ are positive, since they dominated by the 
contribution from the Pauli form factor. Recent experimental evidence 
from the SAMPLE Collaboration gives a positive value of 
$G_M^s=0.37 \pm 0.20 \pm 0.26 \pm 0.07$ at $Q^2=0.1$ (GeV/c)$^2$.  
The other experimental values of $G_E^s$ and $G_M^s$ in Figs.~\ref{GEs} 
and \ref{GMs} for $0.4 < Q^2 < 1.0$ (GeV/c)$^2$ were obtained 
\cite{Pate,Frascati} by combining the (anti)neutrino data from E734 
\cite{Ahrens} with the parity-violating asymmetries from HAPPEX \cite{Aniol04} 
and G0 \cite{Armstrong}. 
The theoretical values are in good overall agreement with the 
experimental ones for the entire range $0 < Q^2 < 1.0$ (GeV/c)$^2$. 

\begin{table}
\centering
\caption[]{\small 
Comparison between theoretical and experimental values of 
strange form factors $G_E^s + \eta G_M^s$.}
\label{HappexA4}
\vspace{15pt}
\begin{tabular}{ccccc}
\hline
$Q^2$ & $\eta$ & \multicolumn{2}{c}{$G_E^s + \eta G_M^s$} & \\
(GeV/c)$^2$ & & Present & Experiment & Reference \\
\hline
0.099 & 0.080 & 0.019 & $0.030 \pm 0.028$ & \cite{Aniol05b} \\
0.108 & 0.106 & 0.025 & $0.071 \pm 0.036$ & \cite{Maas2} \\
0.230 & 0.225 & 0.042 & $0.039 \pm 0.034$ & \cite{Maas1} \\
0.477 & 0.392 & 0.047 & $0.014 \pm 0.022$ & \cite{Aniol04} \\
\hline
\end{tabular}
\end{table}

\begin{figure}[htb]
\includegraphics{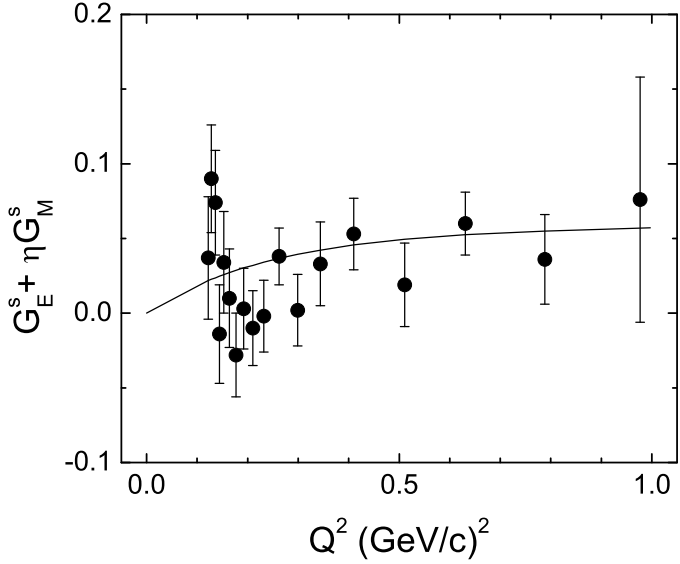} 
\caption[]{\small
Comparison between theoretical and experimental values of 
strange form factors $G_E^s + \eta G_M^s$.  
The experimental values were measured by the G0 Collaboration 
\cite{Armstrong}.}
\label{G0}
\end{figure}

The remaining experimental results of the strange form factors of the 
nucleon obtained by the PVA4, HAPPEX and G0 collaborations correspond 
to linear combinations of electric and magnetic form factors 
$G_E^s+\eta G_M^s$. Table~\ref{HappexA4} and Figure~\ref{G0} show a good 
agreement between the results from the present calculations and the 
experimental data.

\begin{figure}[htb]
\includegraphics{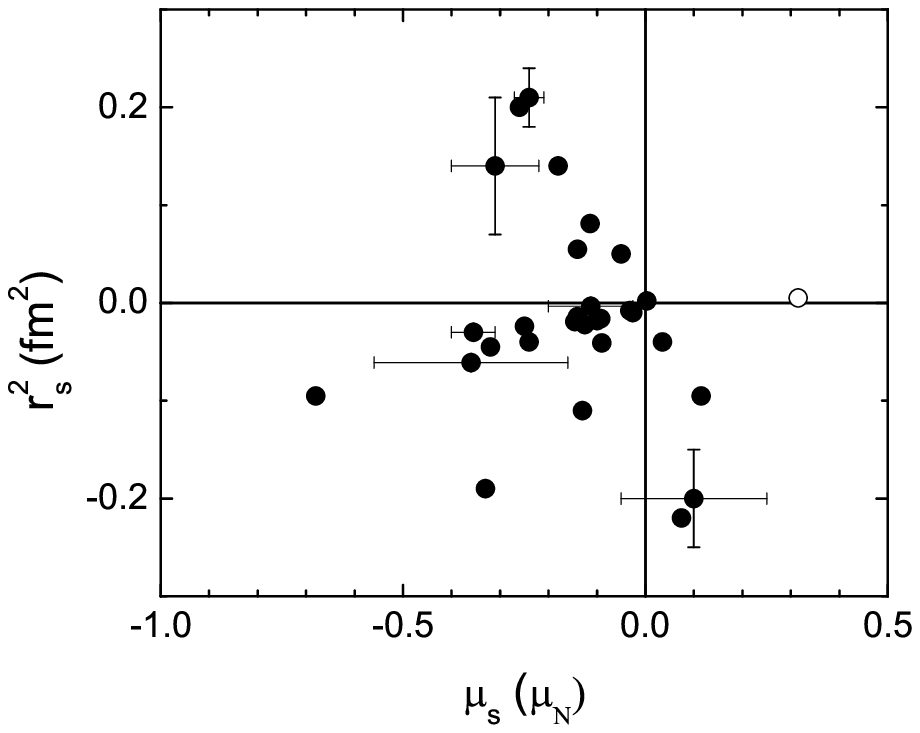} 
\caption[]{\small
Theoretical values of the strange magnetic moments and the 
strangeness radii (filled circles) \cite{beck}. 
The present value is denoted by an open circle.}
\label{mmrad}
\end{figure}

In the majority of theoretical analyses, the strangeness contribution to the 
nucleon is discussed in terms of the static properties, the strange magnetic 
moment $\mu_s$ and the strangeness radius $\left< r_s^2 \right>$. 
Figure~\ref{mmrad} shows a compilation of theoretical values of these 
two quantities (filled circles). Most studies agree on a small negative 
strangeness radius and a moderate negative strange magnetic moment \cite{beck}, 
whereas the results of a combined fit of the strange electric and magnetic 
form factors measured by SAMPLE, PVA4 and HAPPEX at $Q^2 \sim 0.1$ (GeV/c)$^2$, 
$G_M^s(0.1)=0.55 \pm 0.28$ and $G_E^s(0.1)=-0.01 \pm 0.03$ \cite{Aniol05b},  
indicate the opposite sign for both $\mu_s$ and $\left< r_s^2 \right>$. 
Recent lattice calculations give a slightly negative values of the strange 
magnetic moment $\mu_s= -0.046 \pm 0.019$ $\mu_N$ \cite{Leinweber1} and 
the strange electric form factor $G_E^s(0.1)=-0.009 \pm 0.028$ \cite{Leinweber2}. 

In the present approach, the strange magnetic moment is given by
\ba
\mu_s = G_{M}^{s}(0) = \frac{1}{2} (\alpha_{\omega}^s+\alpha_{\phi}^{s}) 
= 0.315 \, \mu_N . 
\label{mus}
\ea
$\mu_s$ does not depend on the mixing angle $\epsilon$, since according to 
(\ref{strange}) one has $\alpha_{\omega}^s+\alpha_{\phi}^{s} = \kappa g_2 \sin \eta_2$. 
Its sign is determined by the sign of the tensor coupling $\alpha_{\phi}^s$ 
($\gg \alpha_{\omega}^s$, see Table~{\ref{fit}). 
The strangeness charge radius is given by 
\ba
\left< r^2_s \right>_E &=& 
-6 \left. \frac{d G_E^s(Q^2)}{d Q^2} \right|_{Q^2=0}
\nonumber\\  
&=& 3 \beta_{\phi}^{s} \left( \frac{1}{m_{\phi}^2}-\frac{1}{m_{\omega}^2} \right) 
+\frac{3}{4M_N^2}(\alpha_{\omega}^{s}+\alpha_{\phi}^{s}) 
= 0.005 \mbox{ fm}^2 .
\label{rsqe}
\ea
The first term is entirely due to the VMD contribution. In the absence of mixing, 
$\beta_{\phi}^s=0$ and the strangeness radius depends on the second term only  
$\left< r^2_s \right>_E = 3\mu_s/2M_N^2 = 0.021$ fm$^2$.
Similarly, the strangeness magnetic radius is given by 
\ba
\left< r^2_s \right>_M &=& 
-\frac{6}{\mu_s} \left. \frac{d G_M^s(Q^2)}{d Q^2} \right|_{Q^2=0}
\nonumber\\  
&=& 6 \left[ 2\gamma + \frac{\beta_{\phi}^{s}+\alpha_{\phi}^{s}}
{\alpha_{\omega}^s+\alpha_{\phi}^{s}} \frac{1}{m_{\phi}^2}
+ \frac{\beta_{\omega}^{s}+\alpha_{\omega}^{s}}
{\alpha_{\omega}^s+\alpha_{\phi}^{s}} \frac{1}{m_{\omega}^2} \right] 
= 0.410 \mbox{ fm}^2 .
\label{rsqm}
\ea
The first term is due to the intrinsic form factor, whereas the last two 
terms arise from the VMD contribution. In the absence of mixing, 
$\beta_{\phi}^s=\beta_{\omega}^s=\alpha_{\omega}^s=0$ and the strangeness  
magnetic radius of (\ref{rsqm}) reduces to two terms only which contribute 
almost the same amount to the radius 
$\left< r^2_s \right>_M = 6[2\gamma+1/m_{\phi}^2] = 0.403$ fm$^2$.

The values of the strangeness contribution to the magnetic moment and the 
charge radius in the two-component model are indicated in Figure~\ref{mmrad} 
by an open circle. The signs of both quantities are found to be positive, 
in agreement with the available experimental evidence. 
A positive value of the strange magnetic moment seems to 
preclude an interpretation in terms of a $uuds\bar{s}$ fluctuation into a 
$\Lambda K$ configuration \cite{Riska}. On the other hand, an analysis of 
the magnetic moment of $uuds\bar{s}$ pentaquark configurations belonging 
to the antidecuplet gives a positive strangeness contribution for states with 
angular momentum and parity $J^P=1/2^+$, $1/2^-$, and negative for $3/2^+$ 
states \cite{BGS}.  

\section{Summary and conclusions}

In summary, in this Letter it was shown that the recent experimental 
data on the strange nucleon form factor can be explained very well in 
a two-component model of the nucleon consisting of an intrinsic 
three-quark structure with a spatial extent of $\sim 0.49$ fm 
surrounded by a meson cloud. The present approach is a combination 
of the two-component model of Bijker and Iachello for the 
electromagnetic nucleon form factors \cite{BI} and a mechanism 
to determine the strangeness content via the coupling of the strange current 
to the $\phi$ and $\omega$ mesons according to Jaffe \cite{Jaffe}. 
The condition that the strange quarks do not contribute to the 
electric charge of the nucleon, reduces the number of independent 
coefficients of the two-component model of \cite{BI} by one. 
The parameters are completely determined by the electromagnetic 
form factors of the proton and neutron. 

The good overall agreement between the theoretical and experimental 
values for the electromagnetic form factors of the proton and neutron 
and their strange quark content shows that the two-componet model 
provides a simultaneous and consistent description of the electromagnetic 
and weak vector form factors of the nucleon. 
In particular, the strange magnetic moment was found to be positive, 
in contrast with most theoretical studies, but in agreement with 
the results from parity-violating electron scattering experiments. 

The first results from the SAMPLE, PVA4, HAPPEX and G0 collaborations 
have shown evidence for a nonvanishing strange quark contribution to 
the charge and magnetization distributions of the nucleon. Future 
experiments on parity-violating electron scattering and neutrino 
scattering hold great promise to make it possible to unravel the contributions 
of the different quark flavours to the electric, magnetic and axial form 
factors, and thus to give new insight into the complex internal structure 
of the nucleon. 

\ack{
This work was supported in part by a grant from CONACYT, Mexico. 
It is a pleasure to thank Franco Iachello for interesting discussions.}

\end{document}